# MIRRORED LANGUAGE STRUCTURE AND INNATE LOGIC OF THE HUMAN BRAIN

## As A Computable Model Of The Oracle Turing Machine

Han Xiao Wen
Weimingbosi Corporation
PKU Biocity No. 39 Shang Di Xi Lu, Haidian
Beijing, 100085 China

We wish to present a mirrored language structure (MLS) and four logic rules determined by this structure for the model of a computable Oracle Turing machine. MLS has novel features that are of considerable biological and computational significance. It suggests an algorithm of *relation learning and recognition* (RLR) that enables the deterministic computers to simulate the mechanism of the Oracle Turing machine, or P = NP in a mathematical term.

A concept of mirrored language structure for the human brain has already been proposed by Chomsky [4] as Universal Grammar (UG). His model consists of a hierarchical (deep and surface) dual language structure and a possible set of innate rules. He also proposed the concept that language is the mirror of the mind [3]. His model has been well acknowledged. The challenge that remains is to determine the universal rules between deep and surface language.

A concept of mirrored hierarchical language structure for the Oracle Turing machine was proposed by Turing [11]. Turing's model can be roughly described as follows: A language $L$ consists of two languages: Oracle language $L_o$ and Turing language $L_t$. The member $x$ of $L_t$ can be accepted or rejected correctly by $L_o$ as member $y$ of $L_o$. This model has been only an abstract concept, and has not been implemented due to lack of an efficient means [5, 6].

The present RLR approach is to apply a model of the human brain [10] as a computable model of the Oracle Turing machine. The human brain model has a pair of "mirrored" languages denoted by $L_p = L_c$, where $p$ stands for perceptual and $c$ for conceptual. That is, there exists correspondent relations between the two languages denoted by $L_p \sqsupseteq p = c \in L_c$. In this structure, the member $|c|$ of language $L_c$ is the class of the member $|p|$ of language $L_p$ denoted by $L_p \sqsupseteq |p| \in |c| \in L_c$, where $|p|$ and $|c|$ denote the length of $p$ and $c$, respectively. That is, there also exists member-class relations between the two languages, where the member of the perceptual language is also the member of the members of the conceptual language iteratively, shown as Fig.1:

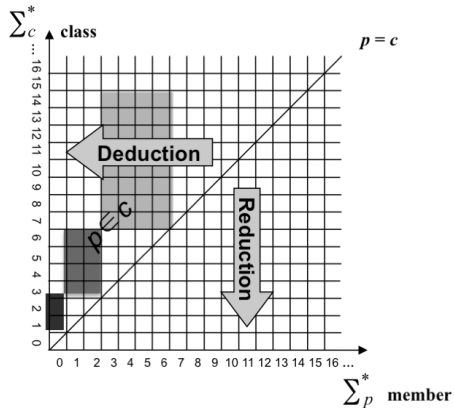

Figure 1. The continuum of member-class relations between perceptual and conceptual language

This mirrored structure embeds four innate *cognitive logic* rules. They can be considered as the universal grammar (UG) that Chomsky has foreseen, and they are specified as follows.

*Sensation*:   Input information is stored in both languages $L_p$ and $L_c$ as correspondent relations denoted by $L_p \sqsupseteq p = c \in L_c$.

*Induction*:   Input relation is stored in between languages $L_p$ and $L_c$ as member-class relations denoted by $L_p \sqsupseteq |p| \in |c| \in L_c$.

*Deduction*:   Output information is retrieved from language $L_p$ to $L_c$ as the class relation mapping denoted by $L_p \geq L_c$.

*Reduction*:   Output information is retrieved from language $L_c$ to $L_p$ as the membership relation mapping denoted by $L_p \leq L_c$.

Formally the model of the human brain is a *relation learning and recognition* language $L_r$ over $\Sigma_r$, $r = p, c$. Let $\Sigma_p$ and $\Sigma_c$ be two identical (mirrored) finite alphabets, and let $\Sigma_p^*$ and $\Sigma_c^*$ be two sets of finite identical strings over $\Sigma_p$ and $\Sigma_c$. Then the language over $\Sigma_p$ is a subset $L_p$ of $\Sigma_p^*$, and the language over $\Sigma_c$ is a subset $L_c$ of $\Sigma_c^*$. Thus $p \in L_p \Leftrightarrow c \in L_c$, iff $L_p \ni p = c \in L_c$ and $L_p \ni |p| \in |c| \in L_c$ for all $p \in \Sigma_p^*$ and $c \in \Sigma_c^*$.

By definition *relation learning and recognition* language $L_r$ is an Oracle Turing machine (nondeterministic Turing machine). *Relation learning and recognition* language $L_r$ is also a deterministic Turing machine, which is defined and specified [6] by:

a countable set of domain $D = \Sigma^*$,
a countable set of range $R = \Sigma^* = \{ACCEPT, REJECT\}$,
a finite alphabet $\Delta$ such that $\Delta^* \wedge R = \phi$,
an encoding function $E: D \rightarrow \Delta^*$,
a transition function $\tau: \Delta^* \rightarrow \Delta^* \cup R$,

such that relation recognition $r(p, c) \Leftrightarrow p \in L_r$ for all $p, c \in \Sigma_r^*$.

It has not escaped my notice that this model immediately suggests an iterative conception of set that was preliminarily described by Gödel [1, 2, 8], based on which Gödel was able to foresee a polynomial algorithm [9] to solve Yes-or-No problems. However there was a missing link between the iterative conception of set and the algorithm of Yes-or-No problems; as Parsons stated, "it is not so clear as it should be what this conception *is*." [7] It *is* the *mirrored language* with *relation learning and recognition* rules that can bridge the gap. Specifically RLR provides a polynomial means of member-class relation storage and a polynomial means of deductive and reductive relation recognition.

I want to express my gratitude to Professor Chen Chou Liang at Peking University, Professor Roger Tarpy at Bucknell University and Professor James Anderson at Brown University for their guidance and faithful support. My work is a biological and mathematical continuation of Dr. Chomsky's Universal Grammar. My deepest gratefulness goes to him.